\documentclass[10pt,a4paper]{article}
\usepackage[utf8]{inputenc}
\usepackage[T1]{fontenc}
\usepackage{graphicx}
\usepackage{amssymb,amsthm,amsmath}
\usepackage{xcolor,hyperref,titlesec,fancyhdr,etoolbox}
\usepackage{lipsum}
\begin{document}
\title{\Large Continuous variable entanglement between propagating optical modes using optomechanics} 
\author{Greeshma Gopinath$^1$, Yong Li$^{2,3}$, Sankar Davuluri$^{1,*}$\\$^1$ Department of Physics, BITS Pilani, Hyderabad Campus,\\ Hyderabad 500078, India\\$^2$ Center for Theoretical Physics and School of Science,\\ Hainan University, Haikou 570228, China\\$^{3}$\textit{yongli@hainanu.edu.cn}\\$^*$\textit{sankar@hyderabad.bits-pilani.ac.in}}
\date{\today}
\maketitle
\begin{abstract}
	This article proposes a new method to entangle two spatially separated output laser fields from an optomechanical cavity with a membrane in the middle. The radiation pressure force coupling is used to modify the correlations between the input and the output field quadratures. Then the laser fields at the optomechanical cavity output are entangled using the quantum back-action nullifying meter technique. The effect of thermal noise on the entanglement is studied. For experimentally feasible parameters, the entanglement between the laser fields survives upto room temperature.
\end{abstract} 

\section{Introduction}
One of the most interesting features of quantum mechanics is entanglement~\cite{Hensen,Lanyon} 
which is a prerequisite for most of the quantum technologies, such as quantum key distribution\cite{Madsen}, quantum information \cite{Braunstein,Kimble2008}, quantum teleportation \cite{Furusawa,PhysRevLett.70.1895} etc. It has been created in several microscopic systems like atoms\cite{Ritter2012,Julian}, ions\cite{Rowe2001,Blatt2008}, mechanical mirror\cite{ling-zhou-19,Huang_2009,Vitali_2007} etc. Entanglement can be created either in localized objects or propagating (also known as flying) objects. The entanglement in flying photons\cite{PhysRevA.93.033842} is interesting for quantum communication \cite{PhysRevLett.69.2881,Gisin2007} as the entangled states are readily propagating in space. While the entanglement in localized oscillators is interesting for quantum memory \cite{PhysRevA.91.032309,Lvovsky2009} application as they can store quantum states in localized oscillators. Along with the entanglement generation, the rate \cite{Deng_2016,PhysRevA.93.033842} at which entangled pair production from a source is also an important parameter for quantum technologies. Instead of producing discrete entangled pairs, entanglement can also be created in continuous variables. In this paper, we propose a new method to generate continuous variable entanglement between two propagating laser fields using optomechanics.

\par An optomechanical cavity (OMC) couples~\cite{kippenberg-om,aspelmeyer-rmp,meystre,vahala} optical degrees of freedom with mechanical degrees of freedom through an oscillating optomechanical mirror (OMM)\cite{Jayich_2008,Thompson2008,Saarinen:23}. The radiation pressure force inside the OMC displaces the OMM which in turn changes the cavity length and the properties of the outfield from the OMC. The quantum nature of radiation pressure force can induce quantum mechanical perturbation into the OMM and vice versa. In particular, OMC has been used to create entanglement between cavity fields\cite{clerk-13,StefanoPirandola_2003,Vgio}, between cavity field and mechanical oscillator\cite{Chen:17,Liu:17,vitaliD,PhysRevA.103.023525,PhysRevA.100.063805,Aoune2022}, and between two mechanical oscillators\cite{ling-zhou-19,Huang_2009,Vitali_2007,amarendra,Amazioug2020,PhysRevA.89.014302}. Entanglement in optomechanical systems has also been studied under several cases like resolved sideband regime\cite{PhysRevA.79.024301}, reservoir engineering \cite{clerk-13, PhysRevA.96.053831}, and pulsed light \cite{PhysRevA.84.052327}. Entanglement is created using three optomechanically coupled modes in resolved sideband regime\cite{PhysRevA.79.024301,PhysRevLett.99.250401}. The effective interaction between the three modes is reduced to beam-splitter and two-mode squeezing interactions by driving on the sidebands. The two-mode squeezing interaction leads to entanglement which can be swapped with third mode by the beam-splitter interaction. Here we propose a new method to create continuous variable entanglement between two flying optical modes using optomechanics. The technique described in this work can create entanglement in both resolved and unresolved sideband regimes. For experimentally feasible parameters\cite{Rossi2018}, the entanglement generated between the optical modes using this technique can survive upto room temperature. Although there are several mathematical conditions \cite{bell,Peres,duan,zubairy,doherty-04,simon,werner} which can evaluate entanglement, we specifically use the Duan criteria\cite{duan} to establish entanglement. 

\section{\label{sec:level2}System}

\begin{figure}[htb]
	\centering
	\includegraphics[]{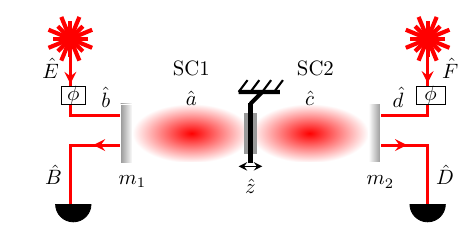}
	\caption{Optomechanical cavity with perfectly reflective mechanical
		mirror in the middle. The mirrors $m_1$ and $m_2$ are rigidly fixed. Phase $\phi$=$\tan^{-1}(-2\Delta/\zeta)$ is added
		such that the input field annihilation operators
		$\hat{b}$ and $\hat{d}$ are related to the laser field annihilation operators $\hat{E}$ and $\hat{F}$ as $\hat{b}=\hat{E}e^{-i\phi}$
		and
		$\hat{d}=\hat{F}e^{-i\phi},$ respectively.}\label{figure2}
\end{figure}
Consider the system shown in Fig.\ref{figure2}. The partially transparent mirrors $m_1$ and $m_2$ are rigidly fixed and they have the same decay rate of $\zeta$. The OMM in the middle is perfectly reflective and its position is given by $\hat{z}$.  The OMM divides the
total cavity into subcavity 1 (SC1) and subcavity 2 (SC2) each with length $l$ and eigenfrequency $\omega_e$. The annihilation operators for the optical field in SC1 and SC2 are given as $\hat{a}$ and $\hat{c}$, respectively. There is no tunneling of $\hat{a}$ into $\hat{c}$ and vice versa, as the OMM is perfectly reflective. The SC1 and SC2 are driven by input fields with annihilation operators $\hat{b}$ and $\hat{d}$, respectively. The output fields from the SC1 and SC2 are represented with annihilation operators $\hat{B}$ and $\hat{D}$, respectively. The Hamiltonian $\hat{H}$ of the total optomechanical cavity is given~\cite{law} as
\begin{eqnarray}
	\hat{H}=\frac{\hbar(\Delta-g\hat{z})}{4}(\hat{x}_a^2+\hat{y}_a^2)+\frac{\hbar(\Delta+g\hat{z})}{4}(\hat{x}_c^2+\hat{y}_c^2)+\frac{\hat{p}^2}{2m}+\frac{m\omega_m^2\hat{z}^2}{2}+\hat{H}_r,\label{eq10}
\end{eqnarray}
where $\hat{p}$,$\omega_m$ and $m$ are momentum, eigenfrequency, and the effective
mass of the mechanical oscillator, respectively. $\hat{x}_o = \hat{o}^\dagger+\hat{o}$, $\hat{y}_o=i(\hat{o}^\dagger-\hat{o})$ with $o = a, c,b,B,d,D$. $\hat{H}_r$ is the Hamiltonian for the environment and its coupling with OMC and $g = \omega_e/l$ \cite{PhysRevLett.109.063601,PhysRevLett.104.083901,PhysRevLett.103.100402} is the optomechanical coupling constant. The detuning $\Delta=\omega_e-\omega_l$
with $\omega_l$ as the frequency of the driving fields.  
\par The dynamics of the optomechanical interaction are given as
\begin{eqnarray}
	\dot{\hat{x}}_a+\dot{\hat{x}}_c=\Delta(\hat{y}_a+\hat{y}_c)-g\hat{z}(\hat{y}_a-\hat{y}_c)-\frac{\zeta}{2}(\hat{x}_a+\hat{x}_c)+\sqrt{\zeta}(\hat{x}_b+\hat{x}_d),\label{eq11}
\end{eqnarray}
\begin{eqnarray}
	\dot{\hat{y}}_a-\dot{\hat{y}}_c=-\Delta(\hat{x}_a-\hat{x}_c)+g\hat{z}(\hat{x}_a+\hat{x}_c)-\frac{\zeta}{2}(\hat{y}_a-\hat{y}_c)+\sqrt{\zeta}(\hat{y}_b-\hat{y}_d),\label{eq12}
\end{eqnarray}
\begin{eqnarray}
	m\ddot{\hat{z}}+\gamma m\dot{\hat{z}}+\omega_m^2 m\hat{z}=\frac{\hbar g}{4}(\hat{x}_a^2+\hat{y}_a^2-\hat{x}_c^2-\hat{y}_c^2)+\hat{\varpi},\label{eq13}
\end{eqnarray}
\begin{eqnarray}\hat{x}_{B}+\hat{x}_{D}=\hat{x}_{b}+\hat{x}_{d}-\sqrt{\zeta}[\hat{x}_a+\hat{x}_c],\label{eq1}\end{eqnarray}
\begin{eqnarray}\hat{y}_B-\hat{y}_D=\hat{y}_b-\hat{y}_d-\sqrt{\zeta}[\hat{y}_a-\hat{y}_c],\end{eqnarray}
where 
$\gamma$ is the decay rate of the OMM and $\hat{\varpi}$ is the thermal noise operator whose correlation is given \cite{vitali-01} as $\langle\hat{\varpi}(\omega)\hat{\varpi}(\omega')\rangle=\hbar m\omega\gamma[1+\coth(\hbar\omega/2k_BT)]\delta(\omega+\omega')$ with $k_B$ the Boltzmann constant, $T$ the
temperature. The operators $\hat{b}$ and $\hat{d}$ are normalized such that their optical powers are given
as $\hbar \omega_l \langle \hat{b}^\dagger \hat{b}\rangle$ and $\hbar \omega_l \langle \hat{d}^\dagger \hat{d}\rangle$, respectively. The left hand sides (LHSs) of Eq.~(\ref{eq11}) and Eq.~(\ref{eq12}) are written as EPR like variables~\cite{duan}, so that the Duan criteria can be verified directly. We linearize Eqs.~(\ref{eq11}-\ref{eq13}) by writing $\hat{O}=\bar{O}+\hat{\delta}_O$, for $O = a,b,c,d,z,B,D$, with $\bar{O}$ being the mean values and $\hat{\delta}_O$ the quantum fluctuation. The steady state position $\bar{z}$ of the OMM can be adjusted by the classical radiation pressure force in SC1 and SC2. By adjusting the optical
fields in SC1 and SC2 such that $\bar{x}_a = \bar{x}_c$ and $\bar{y}_a=\bar{y}_c$, the classical radiation pressure forces on the OMM from SC1 and SC2
are equal and opposite in direction. Then the mean
equilibrium position of OMM is $\bar{z}=0$. We further set the optical fields in SC1 and SC2 to be real by adjusting the phase of the input fields as $\bar{b}=\bar{d}$ = $Ee^{-i\phi}$, where $\phi=\tan^{-1}(-2\Delta/\zeta)$, with $\bar{E}$ being real (see the appendix). Now the equations of motion for the linearized quantum fluctuations are given as 
\begin{equation}
	\dot{\hat{X}}_a+\dot{\hat{X}}_c=\Delta(\hat{Y}_a+\hat{Y}_c)-\frac{\zeta}{2}(\hat{X}_a+\hat{X}_c)+\sqrt{\zeta}(\hat{X}_b+\hat{X}_d),\label{eq14}
\end{equation}
\begin{eqnarray}
	\dot{\hat{Y}}_a-\dot{\hat{Y}}_c=-\Delta(\hat{X}_a-\hat{X}_c)-\frac{\zeta}{2}(\hat{Y}_a-\hat{Y}_c)+\sqrt{\zeta}(\hat{Y}_b-\hat{Y}_d)+2g\bar{x}\hat{\delta}_z,\label{eq15}
\end{eqnarray}
\begin{equation}
	m\ddot{\hat{\delta}}_z+m\gamma\dot{\hat{\delta}}_z+m\omega_m^2\hat{\delta}_z=\frac{\hbar g}{2}\bar{x}(\hat{X}_a-\hat{X}_c)+\hat{\varpi},\label{eq16}
\end{equation}
where $\hat{X}_O = \hat{\delta}_O^\dagger+\hat{\delta}_O$, $\bar{x}_O=\bar{O}^*+\bar{O}$, $\bar{y}_O=i(
\bar{O}^*-\bar{O})$ and $\hat{Y}_O=i(\hat{\delta}_O^\dagger-\hat{\delta}_O)$ with $O = a,c,b,d,B$ and $D$. We have used the relations
$\bar{y}_a=\bar{y}_c=0$, $\bar{x}_a=\bar{x}_c=2\bar{a}:=\bar{x}$ and $\bar{z}$ = 0 in Eqs.~(\ref{eq14}-\ref{eq16}) as we have assumed $\bar{a}=\bar{c}$ and $\bar{a}=\sqrt{\zeta|\bar{E}|^2/(\Delta^2+\zeta^2/4)}$ is real. Achieving $\bar{z}$ = 0
eliminates the classical bistability in the system. Evaluating the Duan criteria requires solving coupled equations Eqs.~(\ref{eq14}-\ref{eq16}), which is intractable analytically in time domain.
Hence we transfer Eqs.~(\ref{eq14}-\ref{eq16}) to the frequency domain by using Fourier transform function $\hat{O}(\omega) = \int_{-\infty}^{\infty} \hat{O}(t) e^{i\omega t}dt/\sqrt{2\pi}$, with $\omega$ the Fourier frequency. 

In frequency domain, Eqs.~(\ref{eq14}-\ref{eq16}) reduce to simple algebraic equations 
\begin{eqnarray}
	\hat{Y}_B(\omega)-\hat{Y}_D(\omega)=e_1[\hat{Y}_b(\omega)-\hat{Y}_d(\omega)]-e_2[\hat{X}_{b}(\omega)-\hat{X}_d(\omega)]+\hat{\varpi}(\omega){e}_3,\label{eq17}
\end{eqnarray}
\begin{eqnarray}
	\hat{X}_{B}(\omega)+\hat{X}_D(\omega)=e_4[\hat{X}_{b}(\omega)+\hat{X}_d(\omega)]-e_5[\hat{Y}_b(\omega)+\hat{Y}_d(\omega)],\label{eq18}
\end{eqnarray}
where
\[
e_1=1+\frac{\zeta(i\omega-\zeta/2)}{(i\omega-\zeta/2)^2-[\alpha(\omega)-\Delta]\Delta},
\quad e_2=\frac{[\alpha(\omega)-\Delta]\zeta}{(i\omega-\zeta/2)^2-[\alpha(\omega)-\Delta]\Delta},\]
\[\centering e_3=\frac{2g\bar{x}\sqrt{\zeta}/[m(\omega_m^2-\omega^2-i\gamma\omega)]}{(i\omega-\zeta/2)(1-\frac{\Delta[\alpha(\omega)-\Delta]}{(i\omega-\zeta/2)^2})},\]\[
e_4=1+\frac{(i\omega-\zeta/2)\zeta}{(i\omega-\zeta/2)^2+\Delta^2},\qquad e_5=\frac{\zeta\Delta}{(i\omega-\zeta/2)^2+\Delta^2},\]
with $\alpha(\omega)=\hbar g^2\bar{x}^2/[m(\omega_m^2-\omega^2-i\gamma\omega)]$. The Eq.~(\ref{eq17}) and Eq.~(\ref{eq18}) are conducive to evaluate the Duan criteria for checking the entanglement between $\hat{B}$ and $\hat{D}$. However the commutation relations for such propagating fields \cite{abram,shepherd-91} are proportional to the Dirac delta distribution. For example, in the
case of input fields, $ [\hat{x}_b(t),\hat{y}_b(t_1)]=2i \delta(t-t_1)$. The divergence arising from the Dirac delta can be eliminated by considering finite time $\tau$ of measurement. This finite time of measurement leads to finite bandwidth $1/\tau$ around the frequency $\omega_f$ at the data is collected. The effect of finite time of measurement can be modeled by
using filter functions $\phi_\tau$ as described in Ref.~\cite{vitali-15}. An important property of $\phi_\tau$ is, for a generic function $f(\omega)$, we can write
\begin{equation}
	\lim_{\tau \to \infty}\int\limits_{-\infty}^{\infty}d\omega\phi_\tau(\omega+\omega_f)\phi_\tau(-\omega-\omega'_f)f(\omega)=\delta_{\omega_f\omega'_f}f(\omega_f),\label{eq24}
\end{equation}
where $\delta_{\omega_f\omega'_f}$ is the Kronecker delta. It is worth emphasizing that $\omega_f$ is constant and its value is chosen according to our needs. We choose $\omega_f$=0 as it corresponds to the output field frequency. Now the relation between the frequency filtered quadrature $\hat{X}_{B_f}$ and $\hat {X}_B$ as
$\hat{X}_{B_f}=\int\limits_{-\infty}^{\infty}d\omega e^{-i\omega t}\phi_\tau(\omega)\hat{X}_{B}(\omega)$. Similarly the frequency filtered quadrature $\hat{Y}_{B_f}=\int\limits_{-\infty}^{\infty}d\omega e^{-i\omega t}\phi_\tau(\omega)\hat{Y}_{B}(\omega)$. Similar relations can be written for the quadratures of $\hat{D}$ also. Note that the frequency filtered quadratures are dimensionless. Then we can write 
\begin{equation}
	\hat{X}_{B_f}(t)+\hat{X}_{D_f}(t)=\int\limits_{-\infty}^{\infty}d\omega e^{-i\omega t}\phi_\tau(\omega)[\hat{X}_{B}(\omega)+\hat{X}_{D}(\omega)],\label{eq25}
\end{equation}
\begin{equation}
	\hat{Y}_{B_f}(t)-\hat{Y}_{D_f}(t)=\int\limits_{-\infty}^{\infty}d\omega e^{-i\omega t}\phi_\tau(\omega)[\hat{Y}_{B}(\omega)-\hat{Y}_{D}(\omega)].\label{eq26}
\end{equation}
By assuming
that $\tau$ is much larger than any characteristic time of system variables, by using Eq.~(\ref{eq17}) and Eq.~(\ref{eq18}) in Eq.~(\ref{eq25}) and Eq.~(\ref{eq26}), the Duan criteria can be evaluated as,
\begin{equation}
	\langle(\hat{X}_{B_f}+\hat{X}_{D_f})^2\rangle+\langle(\hat{Y}_{B_f}-\hat{Y}_{D_f})^2\rangle=[m\gamma k_BT |e_3|^2+2(|ie_1-e_2|^2+|e_4-ie_5|^2) ]|_{\omega=0}.\label{eq27}
\end{equation}
By using the stationary property of the input fields and Eq.~(\ref{eq10}), we can also verify that $[\hat{X}_{B_f},\hat{Y}_{B_f}]=[\hat{X}_{D_f},\hat{Y}_{D_f}]=2i$.

\section{Results}
\begin{figure}[htb]
	\centering
	\includegraphics{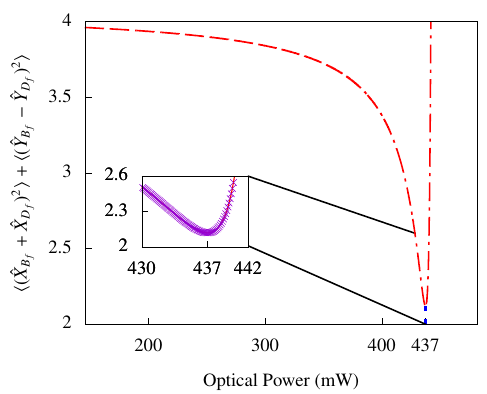}\caption{The Duan criteria as a function of input laser power for $2\Delta/\zeta=10$. Entanglement is created when the Duan criteria is less than four. The inset figure shows the magnified plot of the marked area. Simulation parameters~\cite{Rossi2018}:~$m=2.3\times10^{-12}\,$kg, $\zeta/2\pi= 15.9\times 10^6$\,Hz, $\omega_m/2\pi=1.14\times10^6$\,Hz, $g=4.45\times10^{17}$\,Hz/m, $\omega_l/2\pi=3.77\times10^{14}$\,Hz, $\gamma/2\pi=1.09\times10^{-3}$\,Hz, $T$=300\,K. }\label{plot1}
\end{figure}
The Eq.~(\ref{eq18}) is independent of the optomechanical coupling constant $g$ and hence it can not be influenced by radiation pressure coupling. A straightforward calculation shows that $|e_4-ie_5|^2$ =1 and hence $\langle(\hat{X}_{B_f}+\hat{X}_{D_f})^2\rangle =2$. On the other hand, $g$ enters into Eq.~(\ref{eq17}) through $\alpha(\omega)$. Hence, tweaking the radiation pressure force can influence $\langle(\hat{Y}_{B_f}-\hat{Y}_{D_f})^2\rangle$ term. The presence of $g$  establishes the origin of radiation pressure noise (RPN) in Eq.~(\ref{eq17}). Here we can use RPN suppression techniques to suppress or manipulate the $\hat{X}$ quadrature strength in Eq.~(\ref{eq17}). While there are several techniques\cite{PhysRevLett.99.110801,PhysRevX.2.031016,Moller2017,VYATCHANIN1995269,Vladimir} to suppress RPN, we implement the quantum back-action nullifying meter (QBNM) technique~\cite{davuluri2022} as it allows suppression of quantum back-action when $\omega\to0$. Without going into the details~\cite{Subhash:23} of the RPN suppression, it is simpler to prove entanglement by minimizing $e_1$ and $e_2$. To achieve this we redefine $\alpha(0)-\Delta=u\Delta$ and $\Delta=v\zeta/2$ with $u$ and $v$ being arbitrary real numbers. Then we can write,
\begin{equation}
	\langle(\hat{Y}_{B_f}-\hat{Y}_{D_f})^2\rangle=2|ie_1-e_2|^2=\frac{(1+uv^2)^2+4u^2v^2}{(1-uv^2)^2}.\label{eq28}
\end{equation}
As we already estimated that $\langle(\hat{X}_{B_f}+\hat{X}_{D_f})^2\rangle$=2, $\hat{B}$ and $\hat{D}$ can be entangled only if $\langle(\hat{Y}_{B_f}-\hat{Y}_{D_f})^2\rangle<2$. From Eq.~(\ref{eq28}), such a condition can be satisfied only if $v^2u(1+u)<0$. For $u\not=0$ and $v\not=0$, entanglement can be present only for $-1<u<0$. Minimizing the RHS of Eq.~(\ref{eq28}) further reveals that the smallest possible value for $\langle(\hat{Y}_{B_f}-\hat{Y}_{D_f})^2\rangle$ is
\begin{equation}
	\langle(\hat{Y}_{B_f}-\hat{Y}_{D_f})^2\rangle_{min}=\frac{2}{1+(2\Delta/\zeta)^2},\label{eq29}
\end{equation}
at $\alpha(0)=\Delta[1+(2\Delta/\zeta)^2]/[2+(2\Delta/\zeta)^2]$. As $\alpha(0)$ stems from the radiation pressure force, it is related to the input field power $P$ as,
\begin{equation}
	P=\frac{1+(2\Delta/\zeta)^2}{2+(2\Delta/\zeta)^2}~\frac{m\omega_m^2\Delta\zeta}{16\hbar g^2}((2\Delta/\zeta)^2+1)\hbar \omega_l.\label{neq29} 
\end{equation}
\begin{figure}[htb]
	\centering
	\includegraphics{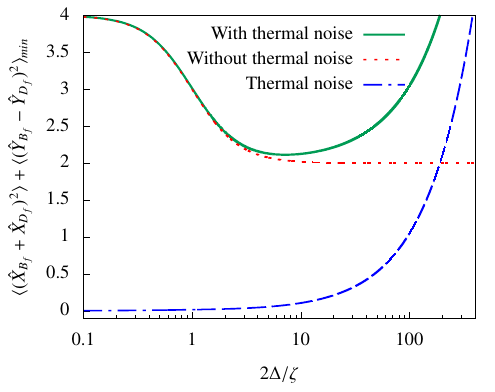}\caption{Minimum value of the Duan criteria as a function of $2\Delta/\zeta$. Entanglement is created when the Duan criteria is less than four. Simulation parameters~\cite{Rossi2018}:~$T=300$\,K, $m=2.3\times10^{-12}\,$kg, $\zeta/2\pi= 15.9\times 10^6$\,Hz, $\omega_m/2\pi=1.14\times10^6$\,Hz, $g=4.45\times10^{17}$\,Hz/m, $\omega_l/2\pi=3.77\times10^{14}$\,Hz, $\gamma/2\pi=1.09\times10^{-3}$\,Hz.}\label{plot2}
\end{figure}
As $2\Delta/\zeta\to \infty$, the RHS of Eq.~(\ref{eq29}) goes to zero. Hence, for the technique described in this article, two is the achievable lower bound for the Duan criteria [or Eq.~(\ref{eq27})]. Note that as $2\Delta/\zeta\to \infty$, $\alpha(0)$ will be equal to $\Delta$, which is the same condition required for QBNM. The entanglement condition $\alpha(0)=\Delta[1+(2\Delta/\zeta)^2]/[2+(2\Delta/\zeta)^2]$, leads to cross-correlations between $\hat{Y}_B$ and $\hat{Y}_D$ quadratures of the output field such that $\langle(\hat{Y}_{B_f}-\hat{Y}_{D_f})^2\rangle\to 0$. 
Thus the radiation pressure coupling leads to cross-correlations which entangle the output fields from SC1 and SC2. By using Eq.~(\ref{eq27}), Fig.~\ref{plot1} is plotted with the Duan criteria on vertical axis and input laser power $P$ on the horizontal axis for $2\Delta/\zeta$ equal to 10. The Duan criteria has minimum value exactly when the optical power is equal to Eq.~(\ref{neq29}). Similarly, the lowest value of the Duan criteria for the curve is given according to Eq.~(\ref{eq29}). The thick green curve in Fig.~\ref{plot2} shows the minimum possible value of the Duan criteria as a function of $2\Delta/\zeta$. The optical power in Fig.~\ref{plot2} is given according to Eq.~(\ref{neq29}) for each point of $2\Delta/\zeta$ . The red dotted curve in Fig.~\ref{plot2} shows the Duan criteria without thermal noise. The blue dashed curve represents the thermal noise (at $T $=300 K) contribution to the Duan criteria. According to the thick green curve, there is no entanglement when  $2\Delta/\zeta$ = 0 and the entanglement increases with increase in $2\Delta/\zeta$ gradually. However, increasing $2\Delta/\zeta$ indefinitely leads to increase in thermal noise which destroys the entanglement. The effect of thermal noise on entanglement is studied in the next section.

\subsection{Temperature Dependence}
\begin{figure}[htb]
	\centering
	\includegraphics{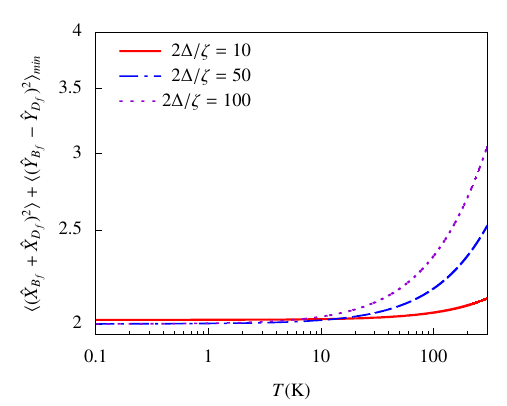}\caption{Minimum value of the Duan criteria as a function of temperature $T$. Entanglement is created when the Duan criteria is less than four. The plot shows entanglement up to 1000\,K, but this may be practically impossible since it can lead to degradation of the OMM, which in turn destroy the entanglement. Simulation parameters~\cite{Rossi2018}:~$m=2.3\times10^{-12}\,$kg, $\zeta/2\pi= 15.9\times 10^6$\,Hz, $\omega_m/2\pi=1.14\times10^6$\,Hz, $g=4.45\times10^{17}$\,Hz/m, $\omega_l/2\pi=3.77\times10^{14}$\,Hz, $\gamma/2\pi=1.09\times10^{-3}$\,Hz. }\label{plot3}
\end{figure}

As shown in Fig.~\ref{plot2}, the entanglement survives even at room temperature for some
$2\Delta/\zeta$ values. Thermal robustness of the entanglement is shown in Fig.~\ref{plot3} for different
$2\Delta/\zeta$ values. The vertical and horizontal axes in Fig.~\ref{plot3} represent the minimum value of the Duan criteria and temperature, respectively. Again the optical power is given according to Eq.~(\ref{neq29}) for each $2\Delta/\zeta$ curve. Entanglement is present even at room temperature in Fig.~\ref{plot3}. Hence, the generated entanglement is robust against temperature. The thermal robustness of entanglement can be quatified by rewriting the thermal noise term in Eq. \ref{eq27} as
\begin{equation} |e_3|^2m\gamma k_BT|_{\omega=0}=\frac{8}{2\pi}\frac{\Delta}{\zeta}\frac{k_BT/\hbar}{Qf},\label{eq30}
\end{equation} where $Q=\omega_m/\gamma$ and $f=\omega_m/2\pi.$ It is known that a value far greater than one for the ratio\cite{aspelmeyer-rmp,Davuluri_2016} $k_BT/(\hbar Q f)$ ensures the thermal environment has negligible effect on the
OMM.  Hence the availability of high quality OMM can make the entanglement immune to room temperature noise. Note that Eq.~(\ref{eq27}) proves that large $2\Delta/\zeta$ value minimizes the Duan criteria in absence of thermal noise. On the other hand, large $2\Delta/\zeta$ increase the thermal noise as shown in Eq.~(\ref{eq30}). Luckily, the increase in thermal noise for large $2\Delta/\zeta$ ratio can be compensated by choosing OMM with high quality factor. The ideal
scenario is $k_BT/(\hbar Q f)\ll1\ll2\Delta/\zeta$. Such scenario is achievable for a wide range of
$2\Delta/\zeta$ ratios as shown by the green curve in Fig.~\ref{plot2}.

\section{\label{subsec:levelA} SIMULATION PARAMETERS}
For simulation, we use the following optomechanical parameters~\cite {Rossi2018}:~$m=2.3\times10^{-12}\,$kg, $\zeta/2\pi= 15.9\times 10^6$\,Hz, $\omega_m/2\pi=1.14\times10^6$\,Hz, $g=4.45\times10^{17}$\,Hz/m, $\omega_l/2\pi=3.77\times10^{14}$\,Hz, $\gamma/2\pi=1.09\times10^{-3}$\,Hz.
\section{\label{sec:level5}Conclusion}
A new method for generating entanglement between two spatially separated propagating laser fields is discussed. By using an OMM in the middle of an OMC, the quantum properties of optical field from one subcavity are communicated to the optical field in the other subcavity through the OMM. This leads to quantum correlations between two laser fields exiting the OMC. We derived the necessary and sufficient conditions under which the correlations can lead to entanglement between the two output fields. Then underlying physics for entanglement generation can be attributed to QBNM technique. Robustness of the generated entanglement against temperature is studied using the Duan criteria. For experimentally feasible parameters, entanglement can be generated even at room temperature in both resolved and unresolved sideband regimes.
\section{Acknowledgements}
This work is supported by the National Natural Science Foundation of China (Grants No. 12274107 and No. 12074030).
\appendix
\section{ Derivation of conditions for phase $\phi$}
While writing Eqs.~(\ref{eq11}-\ref{eq13}) we set $\bar{z}=0$ by adjusting the classical radiation pressure force in SC1 and SC2. Here we show quantitatively the conditions that lead to $\bar{z}=0$. By using Eq.~(\ref{eq10}) the dynamics of mean variables are given as
\begin{equation}
	\ddot{\bar{z}}+\gamma\dot{\bar{z}}+\omega_m^2\bar{z}=\frac{\hbar g}{4m}(\bar{x}_a^2+\bar{y}_a^2-\bar{x}_c^2-\bar{y}_c^2),\label{A1}
\end{equation}
\begin{equation}
	\dot{\bar{x}}_a=(\Delta-g\bar{z})\bar{y}_a-\frac{\zeta}{2}\bar{x}_a+\sqrt{\zeta}\bar{x}_b,\label{A2}
\end{equation}
\begin{equation}
	\dot{\bar{y}}_a=-(\Delta-g\bar{z})\bar{x}_a-\frac{\zeta}{2}\bar{y}_a+\sqrt{\zeta}\bar{y}_b.\label{A3}
\end{equation}
From Eq.~(\ref{A1}), $\bar{z}$ is set to zero by imposing the condition $\bar{x}_a-\bar{x}_c=\bar{y}_a=\bar{y}_c=0$. As optical fields in SC1 and SC2 are
independently driven by two separate laser fields, it is possible
to set $\bar{x}_a=\bar{x}_c$.  We can further set that the mean fields in SC1
and SC2 are real, then $\bar{y}_a=\bar{y}_c=0$ and hence Eq.~(\ref{A2}) implies that $\sqrt{\zeta}\bar{x}_a=2\bar{x}_b$ while Eq.~(\ref{A3}) implies that $\sqrt{\zeta}\bar{y}_b=\Delta\bar{x}_a$.
Using these relations, we can finally write	$\zeta\bar{y}_b=2\Delta\bar{x}_b$, and this implies
\begin{equation}
	\phi=\tan^{-1}\big(\frac{-2\Delta}{\zeta}\big).\label{A4}
\end{equation}
One can also apply the same arguments to $\bar{x}_c$ and $\bar{y}_c$ as well.

\end{document}